\documentclass[a4paper, 12pt,psfig]{iopart}   
\usepackage{iopams}
\usepackage{epsf}
\usepackage{euscript}
\usepackage{graphicx}

\begin{document}
\title {Nodal domains on isospectral quantum graphs:
the resolution of isospectrality ?}
\author{Ram Band$^{1}$,  Talia Shapira$^{1}$  and Uzy Smilansky$^{1,2}$}
\address {$^{1}$Department of Physics of Complex Systems,\\ The Weizmann
Institute of Science, Rehovot 76100, Israel\\
$^{2}$ School of Mathematics, Bristol University, Bristol BS83EQ,
UK}

\begin{abstract}
We present and discuss isospectral quantum graphs which are not
isometric. These graphs are the analogues of the isospectral
domains in $\mathbb{R}^{2}$ which were introduced recently in
[1-5] all based on Sunada's construction of isospectral domains
\cite{Sunada}. After presenting some of the properties of these
graphs, we discuss a few examples which support the conjecture
that by counting the nodal domains of the corresponding
eigenfunctions one can resolve the isospectral ambiguity.

\end{abstract}

\section{Introduction}
\label{sec:introduction}   M. Kac's classical paper ``Can one hear
the shape of a drum" \cite {kac}, triggered intensive research in
two complementary aspects of this problem. On the one hand, a
search for systems for which Kac's question is answered in the
affirmative was conducted, and, on the other hand, various
examples of pairs of systems which are isospectral but not
isometric were identified. In the present paper we shall focus our
attention to quantum graphs and in the following lines will review
the subject of isospectrality in this limited context. The
interested reader is referred to [1-11] for a broader view of the
field where spectral inversion and its uniqueness are discussed.

Spectral problems related to graphs emerge in two distinct ways.
In the first, the spectrum of the connectivity (adjacency) matrix
is considered.  It represents a discrete version of the Laplacian,
and for finite graphs, the spectrum is finite. This set of
problems is often referred to as combinatorial graphs. Quantum
(metric) graphs are obtained by associating the standard metric to
the bonds which connect the vertices. The Schr\"odinger operator
consists of the one-dimensional Laplacians on the bonds
complemented by appropriate boundary conditions at the vertices
(see next section). The spectrum of quantum graphs is unbounded,
and it displays many interesting features which made it a
convenient paradigm in the study of quantum chaos \cite{KS}.

Shortly after the appearance of Kac's paper,  M. E. Fisher
published his work ``On hearing the shape of a drum"
\cite{Fisher}, where he addresses isospectrality for the discrete
version of the Laplacian. Since then, the study of isospectral
combinatorial graphs made very impressive progress. In particular,
several methods to construct isospectral yet different graphs were
proposed. A review of this problem can be found in \cite{Brooks}.
In particular, a method which was originally put forward by Sunada
\cite{Sunada} to construct isospectral Laplace-Beltrami operators
on Riemann manifolds, was adapted for the corresponding problem in
the context of combinatorial graphs. Here we shall go one step
further, and show that it can be also adapted for quantum graphs.

The conditions under which the spectral inversion of quantum
graphs is unique were studied previously. In  \cite
{vonBelow,Carlson99} it was shown that in general, the spectrum
 does not determine uniquely the length of the bonds and their
connectivity. However, it was shown in  \cite{gutkinus} that
quantum graphs whose bond lengths are \emph{rationally
independent} ``can be heard" - that is - their spectra determine
uniquely their connectivity matrices and their bond lengths. This
fact follows from the existence of an exact trace formula for
quantum graphs \cite {Roth,KS}. Thus, isospectral pairs of non
congruent graphs, must have rationally dependent bond lengths. The
Sunada method, which is based on constructing the isospectral
domains by concatenating several copies of a given building block,
automatically provides us with graphs with rationally dependent
lengths. An example of a pair of metrically distinct graphs which
share the same spectrum  was already discussed in \cite{gutkinus}.
In a previous report we have shown that all the known isospectral
domains in $\mathbb{R}^{2}$ \cite {Buser,Okada} have corresponding
isospectral pairs of quantum graphs \cite{Tashkent}. Here, we
shall take the subject one step further, and propose that
isospectral graphs can be resolved by counting their nodal
domains. That is, the nodal counts of eigenfunctions belonging to
the same spectral value are not the same.  The idea that nodal
counts resolve isospectrality was suggested in \cite{GSS05} for a
family of isospectral flat tori in 4-d, and was tested
numerically. The present work offers both rigorous and numerical
evidence to substantiate the validity of this conjecture in a few
examples of isospectral graphs.  For these examples the nodal
counts differences ocur on a substantial fraction of the spectrum.

The paper is organized in the following way. For the sake of
completeness we shall give a short review of some elementary
definitions and facts on quantum graphs. We shall then show how
pairs of isospectral domains in $\mathbb{R}^{2}$ can be reduced to
isospectral pairs of quantum graphs, and discuss their spectra and
eigenfunctions. Finally, we shall study the nodal domains of these
eigenfunctions and show that by counting nodal domains one can
resolve the isospectral ambiguity of the graphs presented.

\subsection {A short introduction to quantum graphs}
\label{subsec:graphs}

We consider finite graphs consisting of $V$ {\it vertices}
connected by $B$ {\it bonds}. The $V \times V$ {\it connectivity
matrix} will be denoted by $C_{i,j}$ : $C_{i,j}=r$ when the
vertices $i$ and $j$ are connected by $r$ bonds, and it vanishes
otherwise. The group of bonds which emerge from the vertex $i$
form a ``star" which will be denoted by $S^{(i)}$. The {\it
valency} $v_i$ of a vertex is defined as the cardinality of the
star $S^{(i)}$ and  $v_i = \sum_{j} C_{i,j}$. Vertices with
$v_i=1$ belong to the graph {\it boundary}. The vertices with
$v_i>1$ belong to the graph {\it interior}. The bonds are endowed
with the standard metric, and the coordinates along the bonds $b$
are denoted by $x_{b}$. The length of the bonds will be denoted by
$L_{b}$, and the total length of the graph is
$\mathcal{L}=\sum_{b}L_{b}$.

The domain of the Schr\"odinger operator on the graph is the space
of functions which belong to Sobolev space $H^2(b)$ on each bond b
and at the vertices they are continuous and obey boundary
conditions as is mentioned in (\ref{eq:bc}). The operator is
constructed in the following way. On the bonds, it is identified
as the Laplacian in 1-d $-\frac{{\rm d}^2 \ }{{\rm d} x^2}$. It is
supplemented by boundary conditions on the vertices which ensure
that the resulting operator is self adjoint. We shall consider in
this paper the Neumann and Dirichlet boundary conditions:
\begin{eqnarray}
{\rm Neumann}\ \ & & \ \ \forall i \  \  : \ \ \  \sum_{b\in
S^{(i)}}  \left. \frac{{\rm d}\ \ \ }{{\rm d} x_{b} }\
\psi_{b}(x_{b})\right |_{x_{b}=0} \ = \ 0\ \ , \nonumber \\
{\rm Dirichlet}\ \ & & \ \ \forall i \ \ :  \ \ \ \ \ \ \ \ \
\left. \psi_{b}(x_{b})\right |_{x_{b}=0} =0\ .
 \label{eq:bc}
\end{eqnarray}
The derivatives in (\ref{eq:bc}) are directed out of the vertex
$i$. \emph{Comment}: The Neumann boundary conditions will be
assumed throughout, unless otherwise stated.
 A wave function with a wave number $k$ can be written as
 \begin{equation}
\psi_{b}(x_{b}) = \frac{1}{\sin kL_{b}}\left (\phi_i \sin
k(L_{b}-x_{b}) +\phi_j\sin kx_{b}\right )
 \label{eq:wavefun}
 \end{equation}
where $b$ connects the vertices $i$ and $j$, where the wave
function $\psi_b$  takes the values  $\phi_i$ and $\phi_j$
respectively. The form (\ref {eq:wavefun}) ensures continuity. The
spectrum $\{k_n\}$ and the corresponding eigenfunctions are
determined by substituting (\ref{eq:wavefun}) in (\ref {eq:bc}).
The resulting homogeneous linear equations for the $\phi_i$ are
written as
\begin{equation}
\forall \ 1\le i\le V \ : \ \ \ \sum_{j=1}^B
A_{i,j}(L_1,\cdots,L_B;k)\phi_j\ =\ 0 \ ,
\end{equation}
and a non trivial solution exists when
\begin{equation}
f(L_1,\cdots,L_B;k) \doteq \det A (L_1,\cdots,L_B;k) =0\ .
 \label{eq:secular}
\end{equation}
 The spectrum $\{k_n \}$, which is a discrete, positive
and unbounded sequence is the zero set of the \emph{secular
function} $f(L_1,\cdots,L_B;k)$. The secular functions of the type
(\ref{eq:secular}) have poles on the real $k$ axis, which renders
them rather inconvenient for numerical studies. The secular
function can be easily regularized in various ways, (See e.g.,
\cite{KS,Tashkent}).

It is easy to show that the complete wave function can be written
down in terms of the vertex wave functions at the interior
vertices with $v_i \ge 3$ only. In the sequel we shall denote
their number by $V_{int}$. This reduces the dimension of the
matrix $A$ above from $V$ to $V_{int}$.

The nodal domains of the eigenfunctions (the connected domains
where the wave function is of constant sign), are of two types.
The ones that are confined to a single bond are rather trivial.
Their length is exactly half a wavelength and their number is on
average  $\frac {k \mathcal{L}}{\pi}$. The nodal domains which
extend over several bonds emanating from a single vertex vary in
length and their existence is the reason why counting nodal
domains on graphs is not a trivial task. The number of nodal
domains in a general graph can be written as
\begin{equation}
\hspace{-15mm}  \nu_n =\frac{1}{2}\sum_{i}\sum_{b\in S^{(i)}}
\left \{
  \lfloor\frac{k_n L_{ b }}{\pi}\rfloor+\frac{1}{2} \left( 1-(-1)^{
      \lfloor\frac{k_n L_{b}}{\pi} \rfloor}
    \mathrm{sign}[\phi_i]\mathrm{sign}[\phi_j]
  \right )\right  \}  -B + V \ .
  \label{eq:NNB2}
\end{equation}
where $\lfloor x \rfloor$ stands for the largest integer which is
smaller than $x$, and $\phi_i, \phi_j$ are the values of the
eigenfunction at the vertices connected by the bond $b$
\cite{GSW}. (\ref{eq:NNB2}) holds for the case of an eigenfunction
which does not vanish on any vertex: $\forall i
\,\,\,\phi_i\neq0$.

Recently Schapotschnikow \cite{Schap06} proved that Sturm's
Oscillation Theorem extends to finite tree (loop-less) graphs: the
number of nodal domains of the $n$'th eigenfunction (ordered by
increasing eigenvalues) is $n$. Berkolaiko \cite{Berko06} have
shown that the number of nodal domains is bounded to the interval
$[n-l,n]$ where $l$ is the minimal number of bonds which should be
cut so that the resulting graph is a tree.

Nodal domains can be also defined and counted in an alternative
way which makes use of the vertex wave functions $\{\phi_i\}$ (see
(\ref {eq:wavefun})) exclusively:

 \centerline {{\it A nodal domain consists of a maximal set of
connected interior vertices ($v_i \ge 3$)}}

 \centerline{{\it
 where the vertex wave functions have the same
sign. }}

This definition has to be modified if any of the $\phi_i$
vanishes. Then, the sign attributed to it is chosen to maximize
the number of nodal domains \cite{Leybold}.

We thus have two independent ways to define and count nodal
domains. To distinguish between them we shall refer to the first
as {\it metric} nodal domains, and the number of metric domains in
the $n$'th eigenfunction will be denoted by $\nu_n$. Berkolaiko's
theorem states that $(n-l)\le\nu_n\le n$. The domains  defined in
terms of the vertex wave functions will be referred to as the {\it
discrete} nodal domains. The number of discrete nodal domains of
the $n$'th vertex wave function will be denoted by $\mu_n$. The
sequences of metric and discrete nodal domains counts $\{ \nu_n\}
$ and $\{ \mu_n\} $ are the main objects of study of the present
paper.

\section {Isospectral quantum graphs}
 \label{sec:isospectral}

The first pair of isospectral planar domains which was introduced
by Gordon, Web and Wolpert \cite {Gordon} is a member of a much
larger set which was discussed in \cite {Buser,Okada}. This was
extended in \cite {4russians} to include domains which differ in
the distribution of boundary conditions (Dirichlet or Neumann)
along their boundaries. The common feature of these sets of pairs
of isospectral domains is that they are constructed using the
Sunada method \cite{Sunada}, and they share a few important and
distinctive attributes:
\begin{itemize}
\item The domains are constructed by concatenating an elementary
``building block" in two different prescribed ways to form the two
domains. A building block is joined to another by reflecting along
the common boundary. The shape of the building block is
constrained only by symmetry requirements, but otherwise it is
quite general.

\item The eigenfunctions corresponding to the same eigenvalue are
related to each other by a \emph{transplantation}. That is, the
eigenfunction in a building block of one domain can be expressed
as a linear combination of the eigenfunction in several building
blocks in the other domain. The transplantation matrix is
independent of the considered eigenvalue.

\item  The construction of these pairs reflects an abstract
algebraic structure which was identified by Sunada \cite{Sunada}.
\end{itemize}
\begin{figure}[h]
\centering \scalebox{.8}{\includegraphics{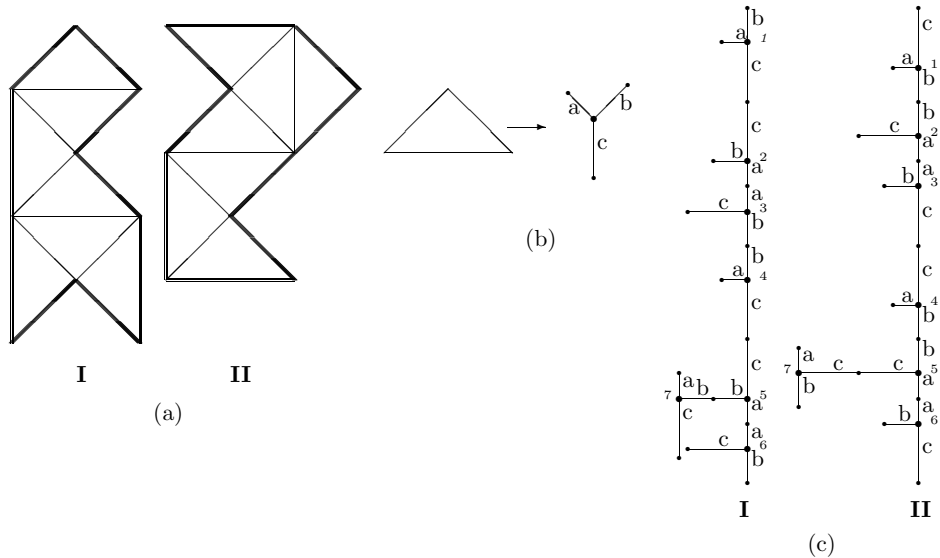}}
 \caption{(a) Planar isospectral domains of the $7_3$ type.
(b) Reducing the building block to a 3-star.(c) The resulting
isospectral quantum graphs.}  \label{fig:7-3}
\end{figure}

An example of an isospectral pair of domains in $\mathbb{R}^2$ and
its building block is given in figure (\ref{fig:7-3}(a)). This is
a pair whose construction is described in \cite{Buser}.  It is
denoted there by $7_3$. Other examples are displayed in e.g.,
\cite {chapman,Buser,4russians}.

We can construct  \emph{metric} graphs which are analogous to
these domains, by replacing the building blocks by  appropriate
graphs which preserve  the requires symmetry. As an example, the
triangular building block in figure (\ref{fig:7-3}(a)) can be
replaced by a 3 - star with bonds of lengths $a$, $b$ and $c$ as
shown in figure (\ref{fig:7-3}(b)). This yields the pair of
isospectral but non isometric graphs shown in figure
(\ref{fig:7-3}(c)). (In drawing figure (\ref{fig:7-3}(c)) we took
advantage of the fact that the ``angles" between the bonds are
insignificant). Note that the two graphs  share the same
connectivity matrix (they are topologically congruent) however
they are not isometric. As a matter of fact the right and the left
graphs are interchanged when the bonds ``$b$" with ``$c$" are
switched. Thus, the lengths $b$ and $c$ must be different to
ensure that the two graphs are not isometric. This is an example
of an asymmetry requirement which the building block must satisfy.
\begin{figure}[h]
\centering
 \scalebox{0.6}{\includegraphics[100,155][460,750]{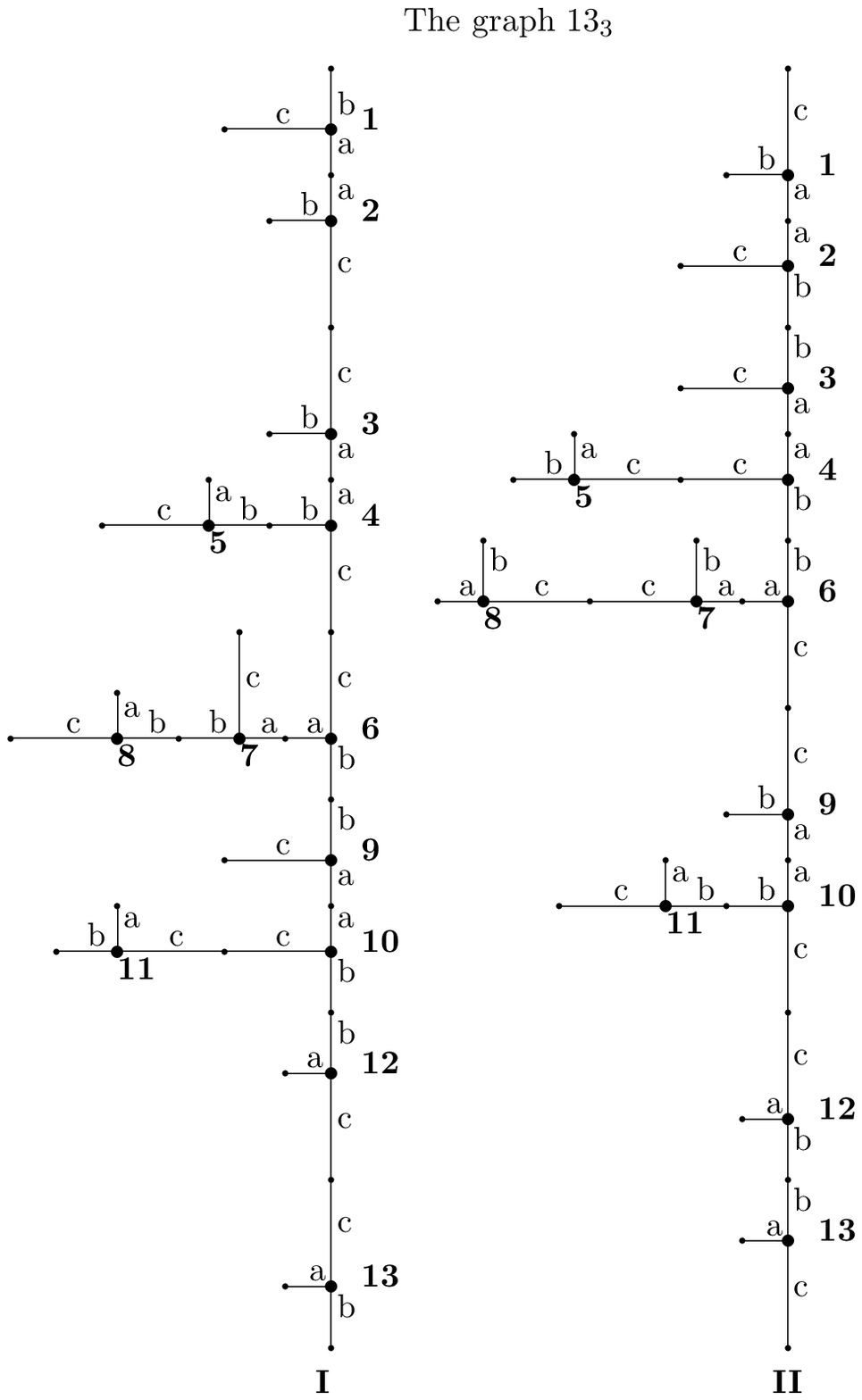}}
 \scalebox{0.6}{\includegraphics[0,153][360,751]{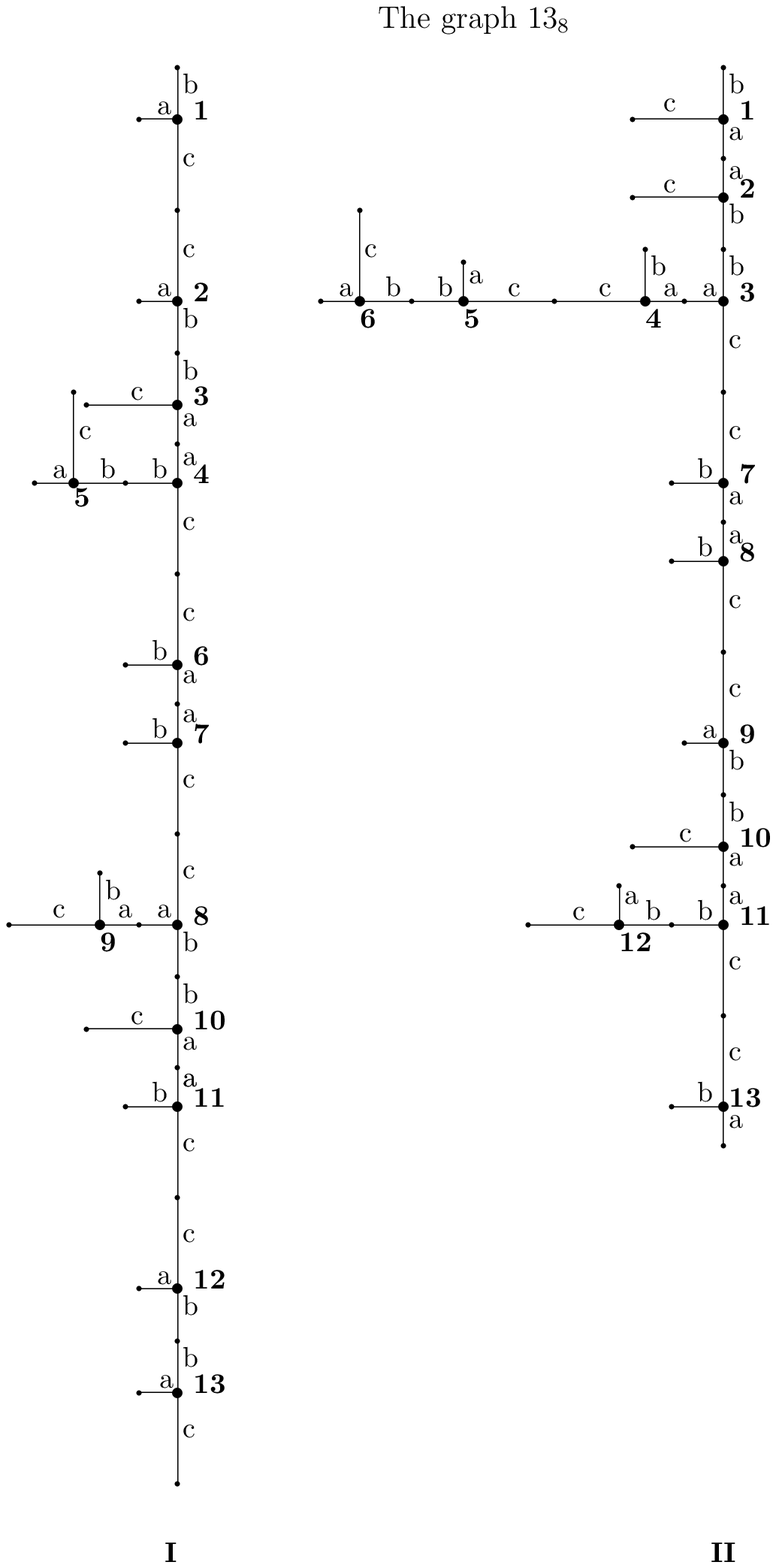}}
  \caption{The
isospectral pairs $13_3$ (left)  and $13_8$ (right) \cite{Buser} }
  \label{fig:13-3and13-8}
\end{figure}
All other pairs of isospectral domains proposed in \cite{Buser}
can be converted to their analogous quantum graphs (see
\cite{Tashkent} for details). Two examples are shown in figure
(\ref {fig:13-3and13-8}).

The most simple pair of isospectral domains can be built by
applying the Sunada method to the Dihedral group $D_4$. The
resulting graphs are shown in figure (\ref{fig:isospectral
pair+notations}), where the letters $D$ or $N$ specify the
boundary conditions at the boundary vertices. (The boundary
conditions at the interior vertices are always Neumann). Note that
the two graphs are connected differently, in contrast with the
previous example where the topology of the two graphs is the same.
In
 \ref {appendix1} we give a brief description of the
construction of this pair, as an example of the Sunada method.

In a way of illustration we shall write explicitly the secular
functions for the graphs shown above. Starting with the $7_3$
pair, we have $V_{int}=7$ (the interior vertices are marked by a
black dot and are enumerated in figure 1(c)). The secular function
for the graph $I$ is written in terms of the matrix
\begin{eqnarray}
\hspace{-2.9cm} \small
 A(a,b,c;k)=\left (
\begin{array}{ccccccc}
\xi - \gamma &     \gamma       & 0          & 0           & 0     & 0     & 0    \\
\gamma      & \xi - \alpha - \gamma & \alpha          & 0          &  0      & 0     &0     \\
0      & \alpha        & \xi - \alpha - \beta   & \beta         & 0      & 0      & 0    \\
 0     & 0       & \beta       &\xi - \beta - \gamma  & \gamma  & 0      & 0    \\
 0     & 0        & 0       & \gamma      & \xi - \alpha - \beta - \gamma& \alpha     &\beta      \\
 0      & 0       & 0       & 0      & \alpha               & \xi - \alpha&
    0   \\
0     & 0        & 0        & 0     & \beta               & 0    &
    \xi - \beta
\end{array}
  \right )
  \label{eq:amatrix}
\end{eqnarray}
Here,
\begin{equation}
\alpha (a;k) = \frac{1}{\sin (2ka)} \ \  ;  \ \ \beta (b;k) =
\frac{1}{\sin (2kb)}\ \ ;\ \ \gamma (a;k) = \frac{1}{\sin (2kc)}
\end{equation}
and
\begin{equation} \xi(a,b,c;k)  =  \tan (ak)\ + \ \tan (bk) \ + \
\tan (ck) \ . \label{eq:star}
\end{equation}
The corresponding vertex eigenfunction $\vec{\phi}^{\ I} =
(\phi^{\ I}_1, \cdots, \phi^{\ I}_7)$  is the eigenvector  of
$A(a,b,c;k_n)$ with a vanishing eigenvalue. The $A$ matrix for the
graph $II$ is obtained from the matrix $A(a,c,b;k_n)$ by
interchanging $c$ and $b$. Explicit computation shows that $\det
A(a,b,c;k)=\det A(a,c,b;k)$. This proves the isospectrality of the
two graphs.

The sum of the elements of any of the columns of $A(a,b,c;k)$ is
$\xi$. Hence the vector $\vec 1 =(1,1,1,1,1,1,1)$ is an
eigenvector with an eigenvalue $\xi$. The values of $k$ for which
$\xi(a,b,c;k)=0$ are in the spectrum of both graphs. They
correspond to eigenfunctions which are the same on each of the
3-stars  - the building block of the complete graphs (figure
1(b)). As a matter of fact, the condition $\xi(a,b,c;k)=0$ is the
secular equation for this 3-star  with bond lengths $a,b,c$ \cite
{KS}. This subset exhausts 1/7 of the spectrum of the graphs, and
in this set the transplantation property is trivial.

The transplantation property which is basic to the proof of
isospectrality for the $\mathbb{R}^2$ domains in \cite{Buser} can
be explicitly formulated  by the transplantation matrix

  {    \small
\begin{eqnarray}
 T=\left (
\begin{array}{ccccccc}
0 & 0  & 0   & 1  & 1  & 1     & 0    \\
0 & 0  & 1   & 0  & 0  & 1     & 1    \\
0 & 1  & 0   & 0  & 1  & 0     & 1    \\
1 & 0  & 0   & 1  & 0  & 0     & 1    \\
1 & 0  & 1   & 0  & 1  & 0     & 0    \\
1 & 1  & 0   & 0  & 0  & 1     & 0    \\
0 & 1  & 1   & 1  & 0  & 0     & 0
\end{array}
\right )\ ,\ \ \ \  {\rm so \ that}\ \ \vec{\phi}^{\ II} = T
\vec{\phi}^{\ I} \ .
\end{eqnarray}     }
We observe that the inverse transplantation is also effected by
$T$: $\vec{\phi}^{\ I} = T \vec{\phi}^{\ II} \ ,$ even though $T$
is not self inversive. The fact that $T$ induces the
transplantation in the two directions, implies that the vertex
wave functions $\vec{\phi}$ must be eigenvectors of $T^2$ (as long
as the eigenvalue is not degenerate). The wave functions are
defined up to normalization, and therefore the corresponding
eigenvalues can be different than unity. The spectrum of $T^2$
consists of the eigenvalue $9$ and the six-fold degenerate
eigenvalue $2$. Thus, the vertex wave functions are either
proportional to $\vec{1}$, or belong to the 6 dimensional subspace
of vectors orthogonal to $\vec{1}$. This observation will be used
in the next section when the nodal structure of the eigenfunctions
is to be discussed.

Other examples of pairs of isospectral graphs such as e.g., the
pair described in \cite{gutkinus} can be obtained from the
isospectral graphs constructed above, by setting the length of the
$c$ bond to zero. This was explicitly shown in \cite{Tashkent}.
Using similar special cases one can generate a rich  variety of
isospectral but not isometric quantum graphs.

Finally, it should be emphasized, that the skeleton of interior
vertices form graphs which are identical to the topological,
colored graphs which were introduced in \cite{Okada} to express
the transplantation properties of the isospectral domains in
$\mathbb{R}^2$. The quantum, metric graphs can be obtained simply
by completing each topological vertex to a 3-vertex star graph
with lengths $a,b,c$. Topological bonds are associated with
lengths according to their colors, and bonds are added to the
topological vertices with valency less than three, to complete
them to 3- stars. This proves the statement made above that the
computation carried out here can be repeated for any of the
topological graphs shown in \cite{Okada}. The graphs corresponding
to the isospectral domains $13_3$ and $13_8$ are shown in figure
\ref {fig:13-3and13-8}.

\begin{figure}[h]
  \centering
  \scalebox{0.5}{\includegraphics[0,0][295,149]{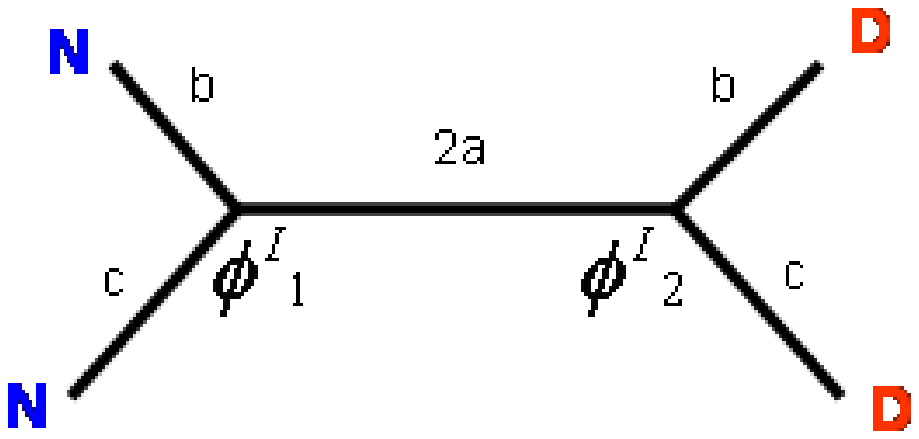}}
  \scalebox{0.5}{\includegraphics[-150,0][330,137]{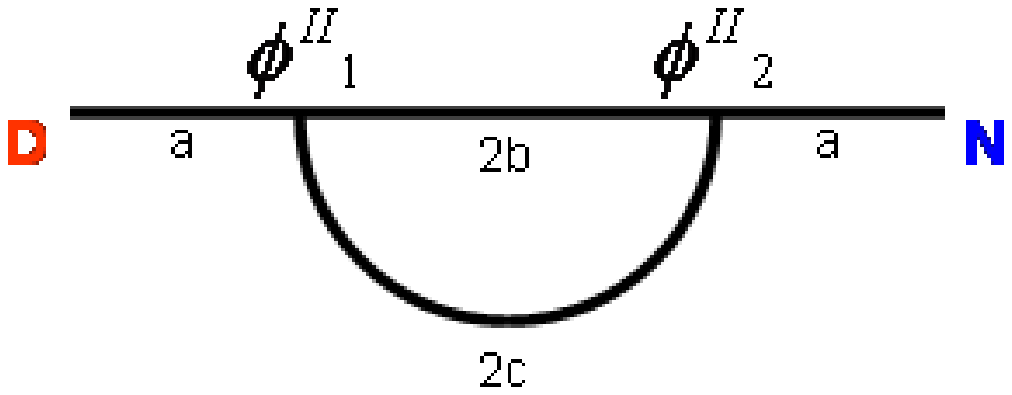}}
 \\   I \hspace{+7.5cm}  II
  \caption{The isospectral pair with boundary conditions. D stands
for Dirichlet and N for Neumann.}
  \label{fig:isospectral pair+notations}
\end{figure}

The Dihedral graphs are much simpler, with $V_{int}=2$. Their
secular determinants can be shown to be identical functions of
$k$. This is an explicit demonstration that the graphs are indeed
isospectral. The secular equation is obtained from
\begin{eqnarray}
\hspace{-15mm}
 A(a,b,c;k)  =  \left( \begin{array}{cc} \eta
-(\beta+\gamma) & -\alpha
\\ -\alpha &
\eta+(\beta+\gamma)
\end{array}\right)
 \nonumber  \\ \hspace{-15mm}
\nonumber \alpha (a;k)   =   \frac{1}{\sin (2ka)} \ \  ;  \ \
\beta (b;k) = \frac{1}{\sin (2kb)}\ \ ;\ \ \gamma (a;k) =
\frac{1}{\sin (2kc)}
 \\  \hspace{-15mm} \eta(a,b,c;k)  =   \cot(2ka)+\cot(2kb)+\cot(2kc)
 \label{eq:secdihedral}
\end{eqnarray}
As can be easily seen, $\det A$ has poles, and to get a regular
secular function we have to multiply it by
$\sin(2ka)\sin(2kb)\sin(2kc)$. It takes the form
\begin{eqnarray}
\nonumber \hspace{-15mm} \ f(a,b,c;k) = \sin(2ka)&&\left(-2 + 2
\cos(2kb)\cos(2kc) - 3\sin(2kb)\sin(2kc)\right) \\
 &&+ 2\cos(2ka)\sin(2kb+2kc)
 \label{eq:secdireg}
\end{eqnarray}

The transplantation matrix for this pair of graphs is derived in
\ref {appendix1}. It reads
\begin{eqnarray}
 T=\frac{1}{\sqrt{2}}\left (
\begin{array}{cc}
\ 1 &\  -1   \\
\ 1 &\ 1
\end{array}
\right )\ ,\ \ \ \  {\rm so \ that}\ \ \vec{\phi}^{\ II} = T
\vec{\phi}^{\ I} \ .
 \label{eq:transdi}
\end{eqnarray}
The eigenvectors $\vec{\phi}^{\ I,II}$ point at the directions
$\theta^{\ I,II} $, with $\tan(\theta^{\ I,II})=\frac{\phi^{\
I,II}_2}{\phi^{\ I,II}_1}$. The transplantation (\ref{eq:transdi})
implies that $\vec {\phi}^{\ II}$ is obtained from $\vec {\phi}^{\
I}$ by a rotation of $\frac{\pi}{4}$ counterclockwise. Direct
substitutions shows that
\begin{equation}
\tan( \theta^{\ I}(k_n)) = g(a,b,c;k_n)\ ,
 \label {eq:tan}
\end{equation}
where,
\begin{equation}
 g(a,b,c;k)=\cos(2ka)+\sin(2ka)\left(\frac{\cos(2kb)-1}
{\sin(2kb)}+\frac{\cos(2kc)-1}{\sin(2kc)}\right)\ .
\label{eq:gfunction}
\end{equation}

The explicit form of the functions $f(a,b,c;k)$ and  $g(a,b,c;k)$
will be used in the discussion of nodal counting in the next
section.

\section {Nodal counts and the resolution of isospectrality}
 \label{sec:nodalcounts}
The real eigenfunctions of Laplace-Beltrami operators on two
dimensional manifolds display usually an intricate sets of nodal
domains - which are the connected domains where the sign of the
eigenfunction is constant \cite {CH}. Sturm's oscillation theorem
for systems in 1-d, and its extension by Courant to any dimension,
establish the connection between the number of nodal domains and
the spectrum: The number of nodal domains $\nu_n$ of the $n$'th
eigenfunction is bounded by $n$. (the eigenfunctions are arranged
by increasing value of their eigenvalues). Courant's theorem for
combinatorial graphs and for quantum graphs were proved in
\cite{Leybold,GSW}, respectively.

Given a pair of isospectral domains. Are the sequences of nodal
counts $\{\nu_n\}_{n=1}^{\infty}$ identical? In other words, can
one use the information stored in the nodal sequences to resolve
isospectrality? This question was recently discussed in the
context of isospectral flat tori in $\mathbb{R}^{d}$ with $d \ge
4$ \cite{GSS05}, and numerical as well as analytical evidence was
brought to substantiate the conjecture that the nodal sequences
resolve isospectrality. In this section we shall provide rigorous
as well as numerical evidence to show that the same is true for
isospectral quantum graphs of the types discussed above. We shall
start by discussing the \emph{discrete} nodal sequences and then
proceed to the \emph{metric} nodal sequences.

\subsection{The discrete nodal sequences}
The {\it discrete} nodal domains were defined as the maximally
connected sets of interior vertices with vertex wave functions of
equal sign. Their number is denoted by $\mu_n$,  and the nodal
sequence is  $\{\mu_n\}_{n=1}^{\infty}$. We shall prove in the
next subsection that for the dihedral graphs half of the entries
in the sequences of discrete nodal counts are different. We are
not able to provide a similar proof for the more complex graphs,
but we shall bring numerical evidence which supports the
conjecture that their sequences of discrete nodal counts are
different.

\subsubsection{The dihedral graphs}
The number of discrete nodal domains for this pair of dihedral
graphs (see figure \ref {fig:isospectral pair+notations}) can be
either one or two, depending on whether the signs of the
components of the (two dimensional) vertex eigenfunctions have the
same sign or not. In other words (see (\ref{eq:tan})), the number
of nodal domains depend on the quadrant in the $(\phi_1,\phi_2)$
plane where the eigenvectors point: $\mu =1$ in the first and the
third quadrants, and $\mu =2$ in the second and the fourth
quadrants:
\begin{equation}
\mu_n = 1 +\frac{1}{2}\left(1 - {\rm sign}(\tan \theta_n)\right) \
.
\end{equation}
The transplantation implies that the eigenvectors $\vec{ \phi
}_n^{\ II}$ are obtained by rotating $\vec{ \phi }_n^{\ I}$ by
$\frac{\pi}{4}$ counterclockwise. Therefore, $\mu_n^I \ne
\mu_n^{II}$ if the transplantation rotates the vectors across the
quadrant borders. In other words,
\begin{equation}
\nonumber \mu^I_n\neq\mu^{II}_n\Longleftrightarrow\tan(\theta_n)
\in\left\{(-1,0)\cup(1,\infty)\right\}
 \label{eq:condition}
\end{equation}
This observation is essential to our discussion since it expresses
the problem of nodal counting in geometrical terms.

It is convenient to construct finite subsequences  $\{\mu^I_n\}$
and $\{\mu^{II}_n\}$ of discrete nodal count of graphs I and II,
restricted to the spectral points in the interval $ 0\le k_n\le K
$.  We denote the number of terms by $N(K)$.  Define
\begin{equation}
P(K) =  \frac{1}{N(K)}\  \sharp\left\{  \ n\le N(K)\ :\
\mu^{I}_n\ne\mu^{II}_n  \right\} \ .
\end{equation}
We shall now prove
\newtheorem{theorem1}{Theorem}
\begin{theorem1}  \label{thm1}
Consider the dihedral graphs I,II discussed above with rationally
independent bond lengths $a,b,c$. Then
\begin{equation}
\lim_{K\rightarrow\infty} P(K) = \frac{1}{2}.
\end{equation}
\end{theorem1}

\noindent \textit{Proof of theorem \ref{thm1}:} The rational
independence of $a,b,c$ implies that the eigenfunctions never
vanish on the inner vertices and hence that the spectrum is
simple.

In order to study $P(K)$ above, we consider the distribution of
the directions of the eigenfunctions $\vec{\phi}_n^I$ in the
spectral interval. Using (\ref{eq:secdireg},\ref{eq:tan},
\ref{eq:gfunction}) we get
\begin{equation} \label{eq:nodal_dist}
\hspace{-25mm} h(x;K)=\langle\delta(x-\tan \theta_n)\rangle_K
=\frac{1}{N(K)} \int_0^{K}{\rm d}k\
\delta(f(a,b,c;k))\left|\frac{{\rm d}f}{{\rm
d}k}\right|\delta(x-g(a,b,c;k)) \label{eq:ndist1}
\end{equation}
In taking the limit $K\rightarrow \infty$ we use
$N(K)=\frac{2a+2b+2c}{\pi}K$ \cite{KS}. Moreover, since $a,b,c$
are assumed to be rationally independent, $k$ creates an ergodic
flow on the 3-torus $\mathrm{T}_3$ spanned by $r=2ka \ {\rm mod}
2\pi ,\quad s= 2kb\ {\rm mod} 2\pi ,\quad t= 2kc\ {\rm mod} 2\pi $
\cite{Bar&Gas}. Ergodicity implies that the  integral over $k$ in
(\ref{eq:ndist1}) may be replaced by an integral over
$\mathrm{T}_3$ leading to
\begin{eqnarray}
\hspace{-25mm} h(x) &=& \frac{\pi}{2(a+b+c)}\frac{1}{\pi^3}
\int_0^{2\pi}dr\int_0^{2\pi}ds\int_0^{2\pi}dt\
\delta(f(r,s,t))\left|\frac{df}{dk}\right|
\delta(x-g(r,s,t))\nonumber \\
\hspace{-25mm}&=&\frac{1}{2(a+b+c)\pi^{2}}\int_0^{2\pi}ds\int_0^{2\pi}dt
\left(\int_0^{\pi}dr\delta(f(r,s,t))\left|\frac{df}{dk}
\right|\delta(x-g(r,s,t))\right.\nonumber \\
\hspace{-25mm} &&\ \ \ \ \ \ \ \ \ \ \ \ \ \ \ \ \ \ \ \ \ \ \ \ \
\ \ \ \ \ \ \  +\ \left.\int_{\pi}^{2\pi}dr\delta(f(r,s,t))
\left|\frac{df}{dk}\right|\delta(x-g(r,s,t))\right)\nonumber\\
\hspace{-25mm}
&=&\frac{1}{2(a+b+c)\pi^{2}}\int_0^{2\pi}ds\int_0^{2\pi}dt
\left[I_1(s,t;x)+I_2(s,t;x) \right] \label{eq:ndist2}
\end{eqnarray}
Now we note that under the transformation $r\ \mapsto\ r'=(r+\pi)\
{\rm mod} 2\pi\quad$ we have
\begin{eqnarray}
\nonumber f(r',s,t)=-f(r,s,t)\\
\nonumber \frac{{\rm d}f}{{\rm d}k}(r',s,t)=-\frac{{\rmd}f}{{\rmd}k}(r,s,t)\\
\nonumber g(r',s,t)=-g(r,s,t)
\end{eqnarray}

Thus, we conclude that
\begin{eqnarray}
\fl \nonumber I_1(s,t;x)=I_2(s,t;-x) &\quad\Rightarrow\quad h(x)=h(-x)\\
\nonumber &\quad\Rightarrow\quad
\int_{-\infty}^{-1}h(x)dx+\int_{0}^{1}h(x)dx=\int_{-1}^{0}h(x)dx+\int_{1}^{\infty}h(x)dx=\frac{1}{2}\\
\nonumber &\quad\Rightarrow\quad
\lim_{K\rightarrow\infty}P(K)=\frac{1}{2} \qquad _{\square}
\end{eqnarray}

\subsubsection{The graphs $7_3$, $13_3$ and $13_8$}

At this stage we are not able to prove the validity of the
conjecture that the nodal sequences resolve the isospectrality of
the graphs $7_3$, $13_3$ and $13_8$. However,  numerical tests
show that for an appreciable fraction of the wave functions, the
nodal counts are distinct. We shall start the discussion by
considering the case $7_3$, for which we can turn the counting
problem into a geometrical problem.

The vertex wave functions which correspond to the same eigenvalues
are related by transplantation:
\begin{equation}
\vec{\phi}^{\ II} = T \vec{\phi}^{\ I} \ \ \ ;\ \ \ \vec{\phi}^{\
I} = T \vec{\phi}^{\ II} \ , \label{eq:transplant}
\end{equation}
where $\vec{\phi}$ stand now for the vertex wave function
restricted to the 7 interior vertices in these graphs. We have
already shown that eigenvectors $\vec \phi^{I,II}$ are either
proportional to $\vec 1$ (the vector with constant entries)- in
which case both have a single nodal domain, and $\delta \mu_n =0$,
or the $\vec \phi^{I,II}$ are orthogonal to $\vec 1$. The 6
eigenvectors of $T^2$ with eigenvalue $2$ span the orthogonal
subspace: three of them correspond to the eigenvalue $+\sqrt2$ of
$T$, and the other three correspond to the eigenvalue $-\sqrt2$ of
$T$. Any vertex wave function $\vec{\phi}^{\ I}$ which is not
proportional to $\vec 1$, can be written as
\begin{equation}
\vec{\phi}^{\ I} = \cos\alpha\  |+\rangle +\sin \alpha\ |-\rangle
\label{eq:nod+}
\end{equation}
where $|\pm\rangle$ stand for arbitrary normalized vectors in the
three dimensional subspaces mention above, and $\alpha \in
[0,2\pi]$. Thus,
\begin{equation}
\vec{\phi}^{\ II} =T \vec{\phi}^{\ I} =\sqrt2 (\cos\alpha \
|+\rangle - \sin \alpha \ |-\rangle) \label{eq:nod-}
\end{equation}
and the difference in nodal numbers comes only through the change
of sign of the linear combination. We are not able to make further
progress beyond this point, and we shall therefore summarize the
numerical findings.

\nonumber \emph{The graph }$7_3$ - Based on the study of
approximately 6600 eigenfunctions we find that the difference
between the discrete nodal counts can take the values $0, \pm2$
only. The nodal counts are different for $\simeq$ 19 $\%$ of the
wave-functions.

\nonumber \emph{The graph }$13_3$ - Based on the study of
approximately 2000 eigenfunctions we find that the difference
between the discrete nodal counts can take the values $0, \pm2$
only. The nodal counts are different for $\simeq$ 22$\%$ of the
wave-functions.

\nonumber \emph{The graph }$13_8$ - Based on the study of
approximately 4700 eigenfunctions we find that the difference
between the discrete nodal counts can take the values $0, \pm2,
\pm4$ only. The nodal counts are different for $\simeq$ 22 $\%$ of
the wave-functions.

This result is certainly encouraging but not sufficient, and we
are trying various options to substantiate it more rigourously.

\subsection{The metric nodal sequences}
The discussion of the metric nodal sequences is facilitated in the
present examples of isospectral graphs since most of them are
trees. Then, by making use of Schapotschnikow's theorem
\cite{Schap06}, the {\it metric} nodal counts $\nu_n =n$.
 This applies in particular to the
isospectral graphs of the types $7_3$, $13_3$ and $13_8$ which are
tree graphs and $\nu^I_n = \nu^{II}_n = n$. The metric counts do
not resolve the isospectrality in these case. However, less
trivially connected graphs can be constructed by replacing the
basic building block. For such graphs Schapotschnikow's theorem
does not apply, and we expect their metric sequences to be
different. Below we shall show that this is true for the pair of
dihedral graph where graph II is not a tree.

Consider the pair of dihedral graphs, and let $\{\nu^{I,II}_n\}$
denote the metric nodal counts of graphs I and and II. Using (\ref
{eq:NNB2}) we express the difference $\nu^{I}_n - \nu^{I,II}_n $
as \begin{eqnarray}  \label{eq:nd_diff}
 \hspace{-23mm} \delta\nu_n&=&
\nu^I_n-\nu^{II}_n \\
\hspace{-23mm}&=& \frac{1}{2}\left[\right.1 -{\rm
sign}(\phi^I_{2,n}\sin(2k_na))+{\rm
sign}(\phi^{II}_{2,n}\sin(2k_nb))  +{\rm
sign}(\phi^{II}_{2,n}\sin(2k_nc))\left.\right]\nonumber
\end{eqnarray}
To obtain (\ref{eq:nd_diff}) a slight modification of (\ref
{eq:NNB2}) was needed: the term in the sum that corresponds to a
bond with boundary vertex that has Dirichlet boundary conditions
should be modified to be $  \lfloor\frac{k_n L_{ b }}{\pi}\rfloor$
instead of $  \lfloor\frac{k_n L_{ b }}{\pi}\rfloor+\frac{1}{2}
\left( 1-(-1)^{
      \lfloor\frac{k_n L_{b}}{\pi} \rfloor}
    \mathrm{sign}[\phi_i]\mathrm{sign}[\phi_j]
  \right )$.
 In addition while deriving (\ref{eq:nd_diff}) we made use
of the identity $\lfloor
\frac{2x}{\pi}\rfloor-2\lfloor\frac{x}{\pi}
\rfloor=\frac{1}{2}(1-{\rm sign}(\sin(2x)))$ and of the freedom to
choose the sign of the first component of the wave-functions to be
positive. A theorem by Berkolaiko \cite{Berko06} guarantees that
$\delta\nu_n$  can only take the values $0$ and $1$. We would like
to compute the distribution of these values on the spectrum. For
this purpose we consider the spectral interval $0<k_n<K$ and study
the function
\begin{equation}
Q(x;K) = \frac{1}{N(K)}\sharp\{n\le N(k): \delta\nu_n=x \}
\end{equation}
and prove the following theorem.
 \newtheorem{theorem2}[theorem1]{Theorem}
\begin{theorem2}  \label{thm2}
Consider the dihedral graphs I,II discussed above with rationally
independent bond lengths $a,b,c$. Then
\begin{equation}
\exists\ \ \ \lim_{K\rightarrow\infty} Q(x,K) =Q(x)\ \  {\rm and}
\ \  Q(0)=Q(1) = \frac{1}{2}.
\end{equation}
\end{theorem2}

\noindent \textit{Proof of theorem \ref{thm2}:} In order to study
the $Q(x,K)$ above we consider the distribution of $\delta\nu_n$,
which is given by the integral
\begin{equation}
\hspace{-25mm} h(x;K)=\langle\delta(x-\delta\nu_n)\rangle_K
=\frac{1}{N(K)}\int_0^{K}{\rm d}k
\delta(f(a,b,c;k))\left|\frac{{\rm d}f}{{\rm
d}k}\right|\delta(x-\delta\nu(a,b,c;k))\ , \label{eq:ndist3}
\end{equation}
where the spectral secular function $f(a,b,c;k)$ is defined in
(\ref{eq:secdireg}) above, and $\delta\nu(a,b,c;k)$ coincides with
$\delta\nu_n$ for $k=k_n$ and can be written explicitly as
\begin{eqnarray}
\fl  &&\delta\nu(a,b,c;k)=\frac{1}{2}\left[1\right. \\
\fl \nonumber
&-&{\rm sign}\left(\cos(2ka)\sin(2ka)+\sin^2(2ka)(\cot(2kb)+\cot(2kc)-\frac{1}{\sin(2kb)}-\frac{1}{\sin(2kc)})\right)\\
\fl \nonumber&+&{\rm sign} \left(\frac{\sin^2(2kb)\sin(2kc)}
{\sin(2kb)+\sin(2kc)}\left(\frac{1}
{\sin(2ka)}+\cot(2ka)+\cot(2kb)+\cot(2kc)\right)\right)\\
\fl \nonumber &+&\left.{\rm sign}\left(\frac{\sin(2kb)\sin^2(2kc)}
{\sin(2kb)+\sin(2kc)}\left(\frac{1}
{\sin(2ka)}+\cot(2ka)+\cot(2kb)+\cot(2kc)\right)\right)\right] \ .
\end{eqnarray}
Following the same route as in the proof of theorem (\ref{thm1})
we take the limit $K\rightarrow \infty$ while making use of the
ergodic theorem, and replace the $k$ integration by an integration
over the 3 - torus with coordinates $r= 2ka \ {\rm mod} 2\pi ,\ s=
2kb\ {\rm mod} 2\pi ,\ {\rm and} \ t= 2kc\ {\rm mod} 2\pi$.
\begin{eqnarray}
\fl \nonumber
h(x)&=\frac{\pi}{2a+2b+2c}\frac{1}{\pi^3}\int_0^{2\pi}{\rm d}
r\int_0^{2\pi}{\rm d}s\int_0^{2\pi}{\rm
d}t\delta(f(r,s,t))\left|\frac{{\rm d}f}{{\rm
d}k}\right|\delta(x-\delta\nu(r,s,t))
\\\fl &=Q(-1)\delta(x-(-1))+Q(0)\delta(x)+Q(1)\delta(x-1)+Q(2)\delta(x-2)
\label{eq:ndist3}
\end{eqnarray}
The last line follows from the fact that $x$ is an integer which
is written as half the sum of four unimodular numbers. Next we
note that under the transformation $r\mapsto r'=(-r)\ {\rm
mod}2\pi, \quad s\mapsto s'=(-s)\ {\rm mod}2\pi, \quad t\mapsto
t'=(-t)\ {\rm mod}2\pi$ we have
\begin{eqnarray}
\nonumber f(r',s',t')=-f(r,s,t)\\
\nonumber \frac{{\rm d}f}{{\rm d}k}(r',s',t')=\frac{{\rmd}f}{{\rmd}k}(r,s,t)\\
\nonumber \delta\nu(r',s',t')=1-\delta\nu(r,s,t)
\end{eqnarray}
Thus, we conclude that $Q(-1)=Q(2) \quad\wedge\quad Q(0)=Q(1)$.
But due to Berkolaiko's theorem \cite{Berko06} $Q(-1)=Q(2)=0$.
Therefore,
\begin{eqnarray}
\fl\nonumber Q(0)=Q(1) = \frac{1}{2}
 \qquad   _{\square}
\end{eqnarray}

\section {Discussion and summary}

There is now a growing amount of evidence that nodal-count
sequences store information on the geometry of the system under
study, which is similar but not equivalent to the information
stored in the spectral sequence. In a recent paper \cite {GKS06}
it was shown that the nodal count sequence for separable Laplace -
Beltrami operators can be expressed in terms of a trace formula
which consists of a smooth (Weyl - like) part, and an oscillatory
part. The smooth part depends on constants which can be derived
from the geometry of the domain, and the oscillatory part depends
on the classical periodic orbits, much in the same way as the
spectral trace formula. However, what we have shown in the present
paper is that the information stored in the two sequences is not
identical: For the isospectral pairs of graphs considered here,
the nodal-count sequences are different in a substantial way. In
this respect, the nodal sequence resolves isospectrality.

Till this work was done, the conjecture that isospectrality is
resolved by counting nodal domains was substantiated by numerical
studies only. Here we presented for the first time a system where
this fact is proved rigorously. The main breakthrough which
enabled the proof was by formulating the counting  problem in a
geometrical setting. We hope that this way will pave the way to
further analytical studies, where more complex systems will be
dealt with.

Finally, we would like to mention a set of open problems which
naturally arise in the present context: Can one find metrically
different domains where the Laplacians have different spectra but
the nodal counting sequences are the same? A positive answer is
provided for domains in $1$ dimension (Sturm) or for tree graphs
(Schapotschnikow). Are there other less trivial examples?

\section {Acknowledgments}
It is  a pleasure to acknowledge S Gnutzmann, M Solomyak and G
Berkolaiko for valuable discussions and comments. We are indebted
to M Sieber for discussions and for sharing with us his notes
which formed the basis for the results discussed in
\ref{appendix1}. The work was supported by the Minerva Center for
non-linear Physics and the Einstein (Minerva) Center at the
Weizmann Institute, and by grants from the GIF (grant
I-808-228.14/2003), and EPSRC (grant GR/T06872/01. US thanks the
School of Mathematics in Bristol for their hospitality and
support.

\appendix
\section{Construction of the isospectral dihedral graphs}  \label{appendix1}

\begin{figure}[h]
  \centering
  \scalebox{0.5}{\includegraphics[0,0][416,384]{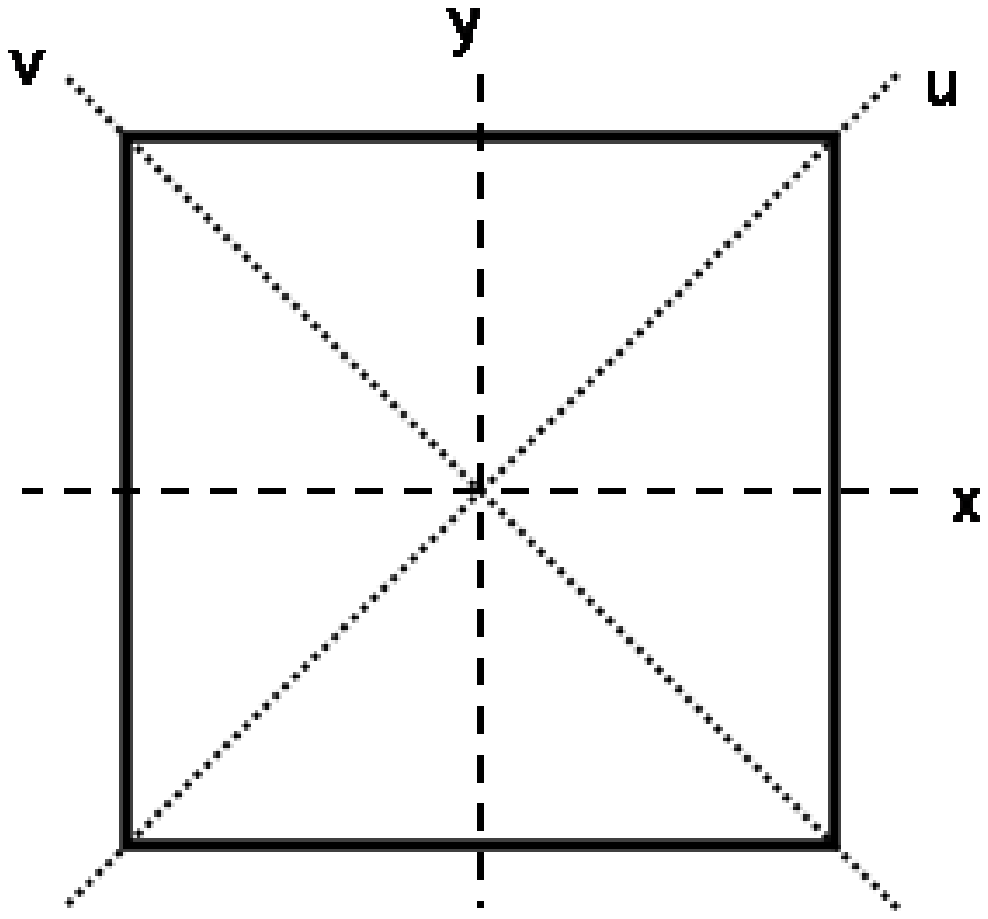}}
  \caption{A square with its 4 possible axis of
  reflection}
  \label{fig:basic square}
\end{figure}

The Dihedral group $\rm D_4 $ is the symmetry group of a square.
Let $a$ be the transformation of rotating the square $\pi/2$
counterclockwise and b be the transformation of reflecting it
along the x axis. The Dihedral group is generated by $a, b$ which
obey $a^{4}=1$,\quad $b^{2}=1,\quad bab=a^{-1}$. The 8 group
elements of ${\rm D}_{4}$ are $\{1,\quad a,\quad a^2,\quad
a^3,\quad r_x=b,\quad r_y=aba^{-1},\quad r_u=ab,\quad r_v=ba \}$.
$ r_x,\, r_y,\, r_u,\, r_v $ are reflections along the axis $x,\,
y,\, u,\, v$ respectively (figure \ref{fig:basic square}). Note
that the multiplication should be read from right to left since
the elements are transformations. This group has 5 irreducible
representations (irreps).  Four of them are one dimensional and
one, which is of our interest, is two dimensional. The two
dimensional representation is:

        \begin{eqnarray}\label{eq:repform1}
        \fl \left\{ \begin{array}{lll}
        1\mapsto\left(\begin{array}{cc}1 & 0 \\ 0 & 1\end{array}\right), &
        a\mapsto\left(\begin{array}{cc}0 & 1 \\ -1 & 0\end{array}\right), &
        a^2\mapsto\left(\begin{array}{cc}-1 & 0 \\ 0 & -1\end{array}\right), \\
        a^3\mapsto\left(\begin{array}{cc}0 & -1 \\ 1 & 0\end{array}\right), &
        r_x\mapsto\left(\begin{array}{cc}-1 & 0 \\ 0 & 1\end{array}\right), &
        r_y\mapsto\left(\begin{array}{cc}1 & 0 \\ 0 & -1\end{array}\right), \\
        r_u\mapsto\left(\begin{array}{cc}0 & 1 \\ 1 & 0\end{array}\right), &
        r_v\mapsto\left(\begin{array}{cc}0 & -1 \\ -1 & 0\end{array}\right)
        \end{array} \right\}
        \end{eqnarray}

Denote the appropriate basis by which the representation looks as
above by $B^{I}=\{\Psi_1^{I}, \Psi_2^{I}\}$. By change of basis
using the matrix $T=\frac{1}{\sqrt{2}}\left(\begin{array}{cc}1 &
-1
\\ 1 & 1\end{array}\right)$ $(x\mapsto T^{-1}xT)$ we obtain the
similar representation
        \begin{equation}\label{eq:repform2}
        \fl \left\{
        \begin{array}{cccc}
        1\mapsto\left(\begin{array}{cc}1 & 0 \\ 0 & 1\end{array}\right), &
        a\mapsto\left(\begin{array}{cc}0 & 1 \\ -1 & 0\end{array}\right), &
        a^2\mapsto\left(\begin{array}{cc}-1 & 0 \\ 0 & -1\end{array}\right), \\
        a^3\mapsto\left(\begin{array}{cc}0 & -1 \\ 1 & 0\end{array}\right), &
        r_x\mapsto\left(\begin{array}{cc}0 & 1 \\ 1 & 0\end{array}\right), &
        r_y\mapsto\left(\begin{array}{cc}0 & -1 \\ -1 & 0\end{array}\right), \\
        r_u\mapsto\left(\begin{array}{cc}1 & 0 \\ 0 & -1\end{array}\right), &
        r_v\mapsto\left(\begin{array}{cc}-1 & 0 \\ 0 & 1\end{array}\right) \end{array} \right\}
        \end{equation}

Denote the second basis by $B^{II}=\{\Psi_1^{II},
\Psi_2^{II}\}$.  So that $\left(\begin{array}{c}\Psi_1^{II} \\
\Psi_2^{II} \end{array}\right) = T\left(\begin{array}{c}\Psi_1^{I}
\\ \Psi_2^{I}
\end{array}\right)$.  Notice that in the first form of
the representation $r_x$ and $r_y$ are diagonal and in the second
one $r_u$ and $r_v$ are diagonal.  This will be exploited soon.

\begin{figure}[!h]
  \centering
  \scalebox{0.5}{\includegraphics[0,0][271,307]{square2a.eps}}
  \scalebox{0.5}{\includegraphics[-100,0][291,326]{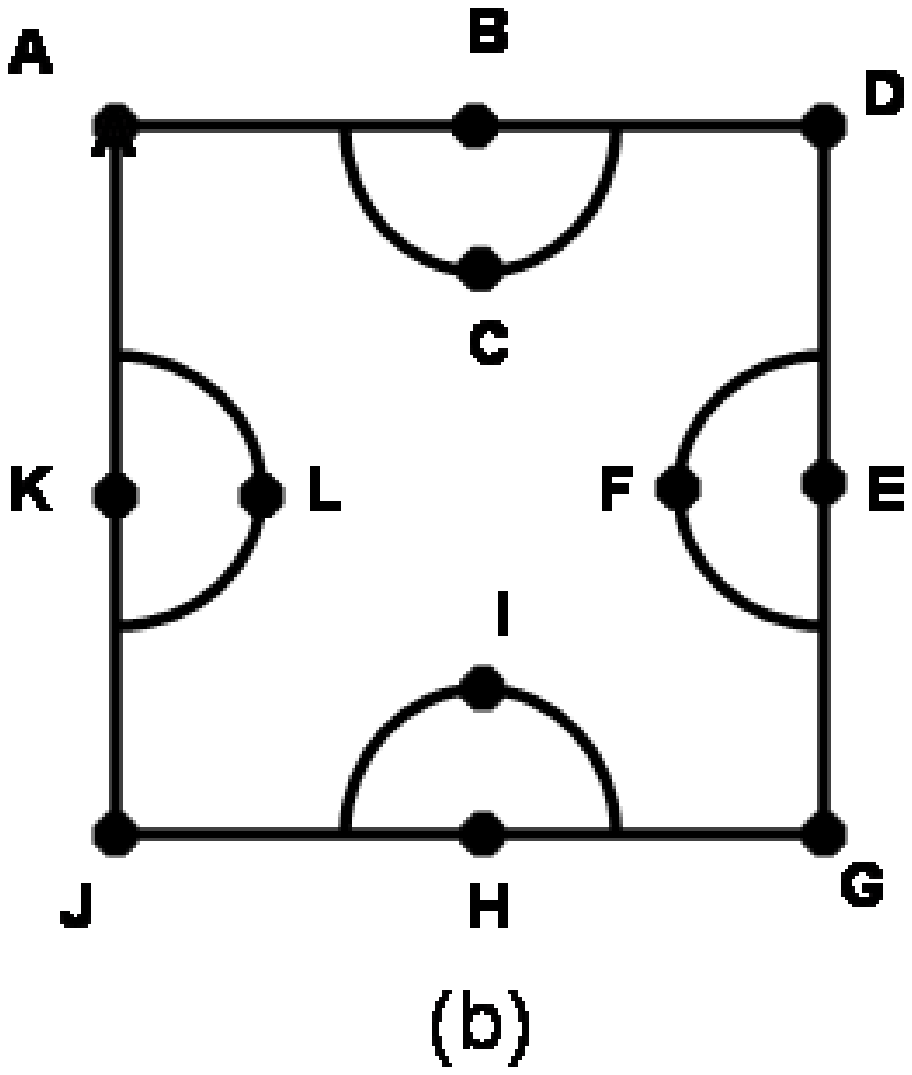}}
  \caption{A graph which obeys the dihedral symmetry.  In (a) the lengths of the bonds are marked.  In (b) The points on which we can deduce boundary conditions are marked.}
  \label{fig:complete graph}
\end{figure}

The graph in figure \ref{fig:complete graph}(a) obeys the dihedral
symmetry. Examine the eigenfunctions of the Schr\"odinger operator
with Neumann boundary conditions on all the vertices.  The set of
all the eigenfuncions of a certain eigenvalue $\lambda$ form a
finite dimensional vector space $V$ with some dimension $dimV=n$.
This vector space is invariant under the action of the dihedral
group and therefore $V$ is the carrier space of some
representation of ${\rm D}_{4}$. We are interested in the case
that this representation is the two dimensional irrep of ${\rm
D}_{4}$ or that this representation is a reducible representation
that contains the two dimensional irrep. Consider the basis
$B^{I}=\{\Psi_1^{I}, \Psi_2^{I}\}$ of this 2d irrep (note that now
$\Psi_1^{I}, \Psi_2^{I}$ have the meaning of wave functions on the
graph in figure \ref{fig:complete graph}). From the matrix which
corresponds to $r_x$ in this basis (see (\ref{eq:repform1})) we
deduce that $\Psi_1^{I}$ vanish on the axis x (as an odd function
with respect to that axis) and as for $\Psi_2^{I}$, its derivative
vanishes on the x axis (since $\Psi_2$ is an even function with
respect to that axis and therefore its derivative is odd). Using
the notation from figure \ref{fig:complete graph}(b), we have:

\begin{eqnarray}
\left.\Psi_1^{I}\right| _{E,F,K,L}=0 \qquad \left. \frac{\rm
d}{{\rm d}x}\Psi_2^{I}\right|_{E,F,K,L}=0 \label{eq:bndcnds1.1}
\end{eqnarray}\\
Similarly, examining the matrix of $r_y$ in this basis
(\ref{eq:repform1}) yields
\begin{eqnarray}
\left.\frac{\rm d}{{\rm d}x}\Psi_1^{I}\right|_{B,C,I,H}=0 \qquad
\left.\Psi_2^{I}\right|_{B,C,I,H}=0 \label{eq:bndcnds1.2}
\end{eqnarray}\\
We can get similar observations by looking on the matrices of
$r_u$, $r_v$, this time for the basis $B^{II}=\{\Psi_1^{I},
\Psi_2^{II}\}$ (\ref{eq:repform2}).  We get the following
\begin{eqnarray}
& \left.\frac{\rm d}{{\rm d}x}\Psi_1^{II}\right|_{D,J}=0 \qquad &
\left.\Psi_2^{II}\right|_{D,J}=0 \label{eq:bndcnds2.1}
\\  &\left.\Psi_1^{II}\right|_{A,G}=0
&\left.\frac{\rm d}{{\rm d}x}\Psi_2^{II}\right|_{A,G}=0
\label{eq:bndcnds2.2}
\end{eqnarray}\\

\begin{figure}[h]
  \centering
  \scalebox{0.5}{\includegraphics[0,0][301,163]{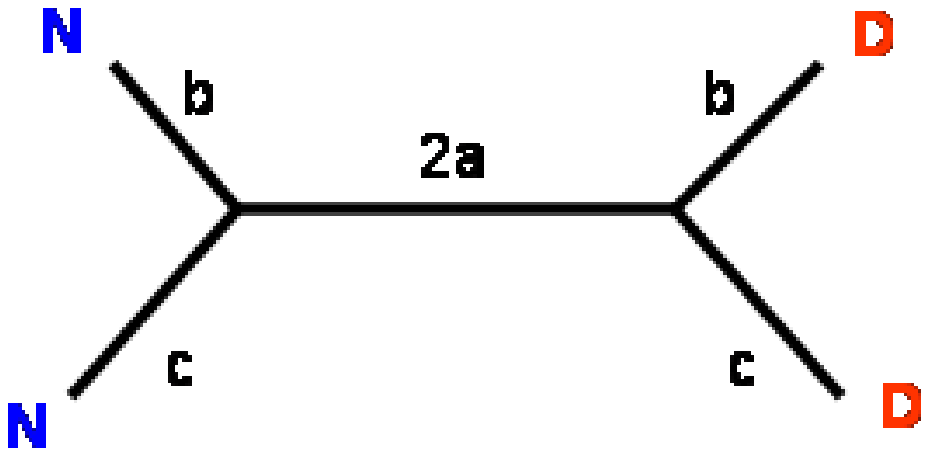}}
  \scalebox{0.5}{\includegraphics[0,0][459,132]{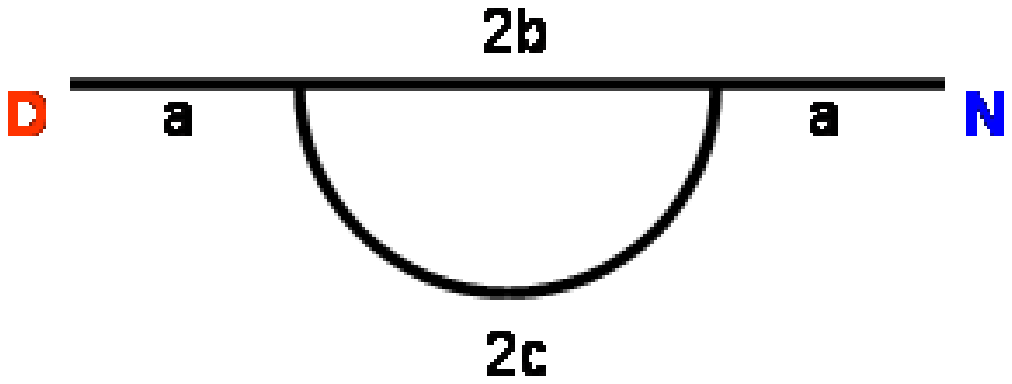}}
  \\   I \hspace{+8cm}  II
  \caption{The isospectral pair with boundary conditions.  D stands for Dirichlet and N for Neumann.}
  \label{fig:boundary conditions}
\end{figure}

Now, a pair of isospectral graphs will be constructed out of the
complete graph.  Each of the graphs in the pair will be a subgraph
of the complete graph. Graph I is the graph that consists of the
vertices B,C,D,E,F and all the bonds which connect them. Graph II
is the graph that consists of the vertices A,B,C,D and all the
bonds which connect them (figure \ref{fig:boundary conditions}).

We can define the subgraphs in another way which will become
convenient later: using the axis $x,y,u,v$ as in figure
\ref{fig:basic square} we define two sets of coordinates
$\left(x,y \right)$ and $\left(u,v\right)$ and use them to examine
the complete graph. Graph I is obtained when restricting ourselves
to the region $D_I\equiv\{x\geqslant0 \quad\wedge\quad
y\geqslant0\}$ of the complete graph. Graph II is obtained when
restricting to the region $D_{II}\equiv\{u\geqslant0
\quad\wedge\quad v\geqslant0\}$. For a certain eigenvalue
$\lambda$ of the complete graph whose representation contains the
2d irrep discussed before, we can define appropriate
eigenfunctions on the two subgraphs by
$\psi^{I}=\left.\Psi_1^{I}\right|_{D_I},
\psi^{II}=\left.\Psi_1^{II}\right|_{D_{II}}$.  The boundary
conditions of $\psi^{I}, \psi^{II}$ are obtained from
(\ref{eq:bndcnds1.1}),(\ref{eq:bndcnds1.2}),(\ref{eq:bndcnds2.1}),
(\ref{eq:bndcnds2.2}) and are shown in figure \ref{fig:boundary
conditions}.

So far we have established the ground for the proof of the
following:
\newtheorem{theorem3}[theorem1]{Theorem}
\begin{theorem3}  \label{thm3}
Let $\sigma^I$ denote the spectra of graph I (with its boundary
conditions) and $\sigma^{II}$ denote the spectra of graph II. Then
$\sigma^I\equiv\sigma^{II}$ with account of multiplicities.
\end{theorem3}
\noindent \textit{Proof of theorem \ref{thm3}:} We will make use
of the results above in order to construct a transplantation which
proves the theorem. Let $\lambda\in\sigma^I$.  Denote by $V^I$ the
space of eigenfunctions which belong to the eigenvalue $\lambda$.
$\forall\psi^{I}\in V^I$ define the functions $\Psi_1^{I},
\Psi_2^{I}$ on the complete graph by the following:
\begin{eqnarray}
\Psi_1^{I}(x,y)=\left\{
\begin{array}{lcl}
\psi^{I}(x,y) & \mbox{for} & x\geqslant 0 \quad\wedge\quad y\geqslant 0 \\
-\psi^{I}(x,-y) & \mbox{for}& x\geqslant 0 \quad\wedge\quad y\leqslant 0 \\
\psi^{I}(-x,y) & \mbox{for} & x\leqslant 0 \quad\wedge\quad y\geqslant 0 \\
-\psi^{I}(-x,-y) & \mbox{for} & x\leqslant 0 \quad\wedge\quad y\leqslant 0 \\
\end{array}\right.\\
\Psi_2^{I}(x,y)=-\Psi_1^{I}(y,-x) \end{eqnarray} Where we have
used again the coordinates $(x,y)$ as defined trivially by the
axis $x,y$ (scaling is obviously not important for our purposes).
It can be verified that $\Psi_1^{I}$, $\Psi_2^{I}$ are valid
eigenfunctions of the eigenvalue $\lambda$ defined on the complete
graph.  If $\Psi_1^{I}(x,y)=c\Psi_2^{I}(x,y)$ for some
$c\in\mathbb{R}$ then we get that $\psi^{I}(x,y)=c\psi^{I}(y,x)
\Rightarrow \psi^{I}\equiv 0$ (this can simply be concluded by
considering the boundary conditions). Therefore $\Psi_1^{I},
\Psi_2^{I}$ are independent eigenfunctions and form a basis
$B^{I}=\{\Psi_1^{I}, \Psi_2^{I} \}$ for the appropriate
eigenspace. This is exactly the basis $B^{I}$ of the first form of
the 2d irrep of $\rm D_4$ (see \ref{eq:repform1}). (This can be
verified for example by examining the representation of the
generators $a,b$ of $\rm D_4$). Now, we construct the basis
$B^{II}=\{\Psi_1^{II}, \Psi_2^{II}\}$ by
$\left(\begin{array}{c}\Psi_1^{II}
\\ \Psi_2^{II}
\end{array}\right) = T\left(\begin{array}{c}\Psi_1^{I}
\\ \Psi_2^{I}
\end{array}\right) = \frac{1}{\sqrt{2}}\left(\begin{array}{cc}1 & -1
\\ 1 & 1\end{array}\right)\left(\begin{array}{c}\Psi_1^{I}
\\ \Psi_2^{I}
\end{array}\right)$.  And now define the function $\psi^{II}$ on
graph II by $\psi^{II}=\left.\Psi_1^{II}\right|_{D_{II}}$.  We get
that $\psi^{II}$ is an eigenfunction of graph II which belongs to
the eigenvalue $\lambda$ and obeys the appropriate boundary
conditions $\Rightarrow \lambda\in\sigma_{II}$.

Following this procedure, one can verify that $T$ is the
transplantation matrix of the vertex wave functions.  We can
similarly describe the inverse transplantation. The inverse
transplantation is linear and therefore if we have an independent
set of eigenfunctions (with the same eigenvalue) they would remain
independent after the transplantation.  Thus, the degeneracies of
$\lambda$ in $\sigma_I$ and $\sigma_{II}$ are equal. The existence
of the inverse transplantation also proves that
$\forall\lambda\in\sigma_{II}\Rightarrow\lambda\in\sigma_I$ and
the degeneracy is again the same.  Therefore
$\sigma_I\equiv\sigma_{II}$ with account of
multiplicities.$\qquad_\square$

\end{document}